%% file: isit.tex
\documentstyle[amssymb,graphicx,graphics,conference]{IEEEtran}

\begin{document}

\renewcommand{\textfraction}{0}

\title{Space-time codes with controllable ML decoding
complexity for any number of transmit antennas}
\author{
\authorblockN{Naresh Sharma}
\authorblockA{
Tata Institute of Fundamental Research \\
Mumbai, India \\
nareshs@ieee.org}
\and
\authorblockN{Pavan R. Pinnamraju}
\authorblockA{
Indian Institute of Technology \\
New Delhi, India \\
pavanramesh.iitd@gmail.com}
\and
\authorblockN{Constantinos B. Papadias}
\authorblockA{
Athens Information Technology \\
Athens, Greece \\
papadias@ait.edu.gr}
}

\date{}

\newcommand{\beq}{\begin{equation}}
\newcommand{\enq}{\end{equation}}
\newcommand{\beqn}{\begin{equation*}}
\newcommand{\enqn}{\end{equation*}}

\newcommand{\beqa}{\begin{eqnarray}}
\newcommand{\enqa}{\end{eqnarray}}
\newcommand{\beqan}{\begin{eqnarray*}}
\newcommand{\enqan}{\end{eqnarray*}}
\newcommand{\beg}{{\bf G}}
\newcommand{\beu}{{\bf Y}}
\newcommand{\bev}{{\bf Z}}
\newcommand{\beom}{{\bf \Omega}}
\newcommand{\bit}{\begin{itemize}}
\newcommand{\eit}{\end{itemize}}
\newcommand{\ca}{{\mathcal{A}}}
\newcommand{\cd}{{\mathcal{D}}}
\newcommand{\ct}{{\mathcal{T}}}
\newcommand{\ch}{{\mathcal{H}}}
\newcommand{\ci}{{\mathcal{I}}}
\newcommand{\crr}{{\mathcal{R}}}
\newcommand{\ce}{{\mathcal{E}}}
\newcommand{\fnorm}{{\mid \mid H \mid \mid^2_F}}
\newcommand{\prob}{\mbox{Prob}}
\newcommand{\sgn}{\mbox{sgn}}
\newcommand{\cn}{{\cal CN}}
\newcommand{\an}{{\phi}}
\newtheorem{lem}{Lemma}
\newtheorem{prop}{Proposition}
\newcommand{\realf}{{\mathbb{R}}}
\newcommand{\compf}{{\mathbb{C}}}
\newcommand{\real}{{\mbox{Re}}}
\newcommand{\imag}{{\mbox{Im}}}
\newcommand{\tr}{{\mbox{Tr}}}
\newcommand{\nt}[1]{{\mbox{\textnormal{#1}}}}
\newcommand{\comment}[1]{}


\newcommand{\beh}{{\bf h}}
\newcommand{\bek}{{\bf K}}
\newcommand{\bet}{{\bf T}}
\newcommand{\bea}{{\bf a}}
\newcommand{\beb}{{\bf b}}
\newcommand{\begn}{{\bf g}}

\maketitle

\thispagestyle{empty}
\def\appendixname{Appendix}

\begin{abstract}

We construct a class of linear space-time block codes
for any number of transmit antennas that have controllable
ML decoding complexity
with a maximum rate of 1 symbol per channel use.
The decoding complexity for $M$ transmit antennas can be varied from
ML decoding of $2^{\lceil \log_2M \rceil -1}$
symbols together to single symbol ML decoding.
For ML decoding of $2^{\lceil \log_2M \rceil - n}$ ($n=1,2,\cdots$)
symbols together, a diversity of $\min(M,2^{\lceil \log_2M \rceil-n+1})$ can
be achieved.
Numerical results show that the performance of the constructed
code when $2^{\lceil \log_2M \rceil-1}$ symbols are decoded together
is quite close to the
performance of ideal rate-1 orthogonal codes (that are non-existent for
more than $2$ transmit antennas).

\end{abstract}

\normalsize

\section{Introduction}

Multiple antenna systems have been of great interest in recent
times, because of their ability to support higher data rates at the same
bandwidth and noise conditions; see e.g.
\cite{vahid},\cite{block2, hott,rajan1} and references therein.
While orthogonal designs offer full diversity with single symbol ML
decoding, they don't have rate $1$ for more than $2$ transmit antennas.

The loss of rate has been addressed by the use of quasi-orthogonal codes
that make the groups of symbols orthogonal where each group has more than
one symbol in general \cite{jafar1,tirkkonen1,pap_fos,np2}. A fully
orthogonal code would have just one symbol per group.
Because of this relaxation of
constraints, these codes achieve higher code rates that were hitherto not
possible with orthogonal codes.
It was shown in \cite{tirkkonen2,np,xia,liu} that performance of above
quasi-orthogonal codes can be improved with constellation rotation.

Codes for any number of transmit antennas were presented in \cite{np2}.
In this paper, we construct that a new class of space-time codes with a
maximum code rate of $1$, that
are inspired from the codes in \cite{np2}, that have a useful property
that the ML decoding is controllable. On one extreme, one can design
rate $1$ codes that have single symbol ML decoding offering diversity
of $2$, and on the other, one can have codes offering full diversity with ML 
decoding of $M/2$ symbols together.

It is, however, shown for the constructed codes that for rate one codes
with single symbol ML decoding, full-diversity is impossible and for codes
that require more than one symbols to be decoded together for ML
symbols decoding, it is indeed possible to have full-diversity.

We use the following notation throughout the paper:
{\small *}, {\small $T$} and {\small $\dagger$} denote the conjugate,
transpose and conjugate transpose respectively of a matrix or a vector;
${\bf I}^M$ and ${\bf 0}^M$ are $M \times M$
identity and null matrices respectively; $\mid \mid A \mid \mid_F$, $\det(A)$
and $\tr(A)$ denote Frobenius norm, determinant and Trace
of matrix $A$ respectively;
$\compf$ denotes the complex number field; $\cn(0,1)$ denotes a circularly
symmetric complex Gaussian variable with zero mean and unit variance.

\section{System Model and Design Criterion}

\label{sysmodel}

Consider a system of $M$ transmit and $N$ receive
antennas. For the ease of presentation, in this paper,
we will assume that $M$ is a power of $2$.
The case of $M$ not being a power of $2$ can be treated easily as in
\cite{np2} by constructing a code of size $2^{\lceil \log_2M \rceil}$
and deleting columns suitably chosen to have the code matrix of size
$2^{\lceil \log_2M \rceil} \times M$.

The statistically independent modulated information symbols
are taken $P$ at a time denoted by
${\bf c} = (c_1,\cdots,c_{P})^T$.
This information vector is pre-coded
(i.e. multiplied) by a $M \times P$ matrix denoted by $\crr$.
Let {\small ${\bf s} = $ $(s_1,\cdots,s_M)^T$} and
\beq
\label{precod}
{\bf s} = \crr {\bf c}
\enq
with $\mbox{E}\{|s_i|^2\} = 1$, $i=1,\cdots,M$.
As we shall soon see, the choice of $\crr$ is central to the
construction of codes.
${\bf s}$ is the input to a linear space-time block
code that outputs a $M \times M$ matrix $G_P[{\bf s}]$, where
\beq
\label{lcode}
G_M[{\bf s}] = \sum_{m=1}^M \left(C_m s_m + D_m s_m^*\right),
\enq
where $C_m$, $D_m$, $m=1,\cdots,M$, are $M \times M$ complex matrices,
which completely specify the code.
This code is transmitted in $M$ channel uses and
the average code rate is hence $P/M$ symbols per channel use.
For a quasi-static fading channel, the received signal is given by
\beq
\label{sigmod}
X[{\bf s}] = \sqrt{\frac{\rho}{M}} G_M[{\bf s}]H + V,
\enq
where $X$ and $V$ are the $M \times N$ received and noise matrices, and $H$
is the $M \times N$ complex channel matrix that is assumed to be constant
over $M$ channel uses and varies independently over the next $M$
channel uses and so on. The entries of $H$ and $V$ are
assumed to be mutually independent and $\cn(0,1)$, and $\rho$ is the
average SNR per received antenna.  We assume that channel is perfectly
known at the receiver but is unknown at the transmitter.

It has been shown in \cite{vahid} by examining the pair-wise probability of
error between two distinct information vectors (say ${\bf u}$, ${\bf v}$
$\in \compf^P$), that for full-diversity, in quasi-static fading channels,
$G_M^\dagger[\crr({\bf u}-{\bf v})]G_M[\crr({\bf u}-{\bf v})]$
should have a rank of $M$. For square code matrices, the above criterion
could be modified to yield
\beq
\label{fulldiv}
\min_{{\bf u},{\bf v}, {\bf u} \neq {\bf v}} \det\{G_M[\crr({\bf u}-{\bf v})]\} \neq 0
\enq

\section{Iterative construction of space-time codes}

The main difference between these codes and those in \cite{np2}
is the choice of $\crr$ that will allow us to vary
the ML decoding complexity and construct full-diversity codes
with decoding of a pair of symbols.

Let us define two disjoint partition vectors that are function of the
vector ${\bf s}$ (whose length will be clear from the context)
denoted by $\ca_{M,1}({\bf s})$
and $\ca_{M,2}({\bf s})$. These partition vectors have same length as
${\bf s}$ and have the same symbols as ${\bf s}$ in indices they possess and
zeros in other indices. If we denote the first and last $M$ elements
of a $2M \times 1$ vector ${\bf s}$ by ${\bf s}_{M,1}$ and ${\bf s}_{M,2}$
respectively, then these partitions are iteratively constructed as
\beqa
\ca_{2M,1}({\bf s}) = \ca_{M,1}({\bf s}_{M,1}) + \ca_{M,2}({\bf s}_{M,2}) \\
\ca_{2M,2}({\bf s}) = \ca_{M,2}({\bf s}_{M,1}) + \ca_{M,1}({\bf s}_{M,2}),
\enqa
and the code is iteratively constructed for the $i$th partition as
\beq
\label{part}
\mbox{$\footnotesize
G_{2M}[\ca_{2M,i}({\bf s})] =
\left[
\begin{array}{rr}
G_M[\ca_{M,i}({\bf s}_{M,1})] & G_M[\ca_{M,\bar{i}}({\bf s}_{M,2})] \\
-G_M[\ca_{M,\bar{i}}({\bf s}^*_{M,2})] & G_M[\ca_{M,i}({\bf s}^*_{M,1})]
\end{array}
\right]$},
\enq
where $\bar{i} = 2$, if $i=1$ and is $1$ otherwise,
and hence by using linearity, we have
\beq
\label{fulliter}
G_{2M}[{\bf s}] = \left[
\begin{array}{rr}
G_M[{\bf s}_{M,1}] & G_M[{\bf s}_{M,2}] \\
-G_M[{\bf s}^*_{M,2}] & G_M[{\bf s}^*_{M,1}]
\end{array}
\right],
\enq
where $G_1[{\bf s}] \stackrel{\Delta}{=} s_1$ $\forall ~ {\bf s} \in \compf^1$,
$\ca_{1,1} = s_1$, and $\ca_{2,1}$ is a null set.

\subsection{Receiver Processing}

We give a practical decoding algorithm to have a low complexity ML
decoding done over a single partition. We note from (\ref{fulliter})
that any row of the constructed code either contains the
symbols ($s_i$'s) or its conjugates (with a possible sign change).
For any ${\bf h} \in \compf^{M\times 1}$, define a transformation denoted by
$\ct$ that takes conjugates of those elements of $M \times 1$ vector
$G_M[{\bf s}]{\bf h}$ that contain conjugates of elements of ${\bf s}$,
and we can write
\beq
\label{equichan}
\ct\{G_M[\ca_{M_i}({\bf s})]{\bf h}\} = \ce_{M,i}({\bf h}) v_{M,i}({\bf s}),
\enq
where $\ce_{M,i}$'s are $M \times (M/2)$ matrices dependent only on ${\bf h}$,
$v_{M,i}$'s are $(M/2) \times 1$ vectors that contain symbols from
partition $i$, with $i=1,2$.
We need a few results from \cite{np2} that we state here without proof.
\ \\ \\
\begin{prop}
\label{prop0}
For any ${\bf h}$, ${\bf s}$ $\in \compf^{M\times 1}$,
\beqa
\nonumber
G_M^\dagger[\ca_{M,1}({\bf s})]G_M[\ca_{M,2}({\bf s})] + ~~~~~~~~~~~~~~~~~~~~~~~~~~ & &\\
~~~~~~~~ G_M^\dagger[\ca_{M,2}({\bf s})] G_M[\ca_{M,1}({\bf s})]  = {\bf 0}^M,
\enqa
\beq
\ce_{M,1}^\dagger({\bf h}) \ce_{M,2}({\bf h}) = {\bf 0}^{M/2},
\enq
\beqa
\nonumber
\mbox{\small $\det\left\{G_{2M}[\ca_{2M,1}({\bf s})]\right\} =
\det\left\{G_{M}[\ca_{M,1}({\bf s}_{M,1}-\hat{{\bf s}}_{M,2})]\right\}$} 
& & \\
\times \mbox{\small $\det\left\{G_{M}[\ca_{M,1}({\bf s}_{M,1}+
\hat{{\bf s}}_{M,2})]\right\}$}, ~~~ & &.
\enqa
where for any $2M \times 1$ vector ${\bf z}$, we define a transformation
denoted by $\hat{\bf z}$ that interchanges the two halves of ${\bf z}$ with
a sign change for the second half, i.e. $\hat{\bf z} = [-z_{M+1},
\cdots, -z_{2M}, z_1, \cdots, z_M]$.
\end{prop}

By taking conjugates appropriately, we can derive a modified signal model
from (\ref{sigmod}) for receive antenna $n$, ($n=1,\cdots,N$), as
\beq
\label{nrx}
\mbox{\small $\hat{X}_n({\bf s}) = \mbox{\large $\sqrt{\frac{\rho}{M}}$}
\left[\ce_{M,1}(H_n) v_{M,1}({\bf s}) +
\ce_{M,2}(H_n) v_{M,2}({\bf s}) \right] + \hat{V}_n$},
\enq
where $H_n$ is the $n$th column of $H$ and $\hat{X}_n$ and $\hat{V}_n$
are derived from $n$th column of $X$ and $V$ respectively by taking the
conjugates of some or all their elements. Let the singular value
decomposition (SVD) of $\ce_{M,i}(H_n)$ be given by
\beq
\label{svdeq}
\ce_{M,i}(H_n) = U_{M,i} S_{M,i} W_{M,i}^\dagger,
\enq
where $U_{M,i}$ and $W_{M,i}$ are unitary
and $S_{M,i}$ is a $M \times (M/2)$ diagonal matrix. Let $\hat{S}_{M,i}$ be a
$M \times (M/2)$ diagonal matrix whose diagonal elements are inverse of
diagonal elements of $S_{M,i}$ and hence
\beq
\label{ssmul}
\mbox{$\hat{S}_{M,i} S_{M,i}^\dagger = \left[
\begin{array}{rr}
{\bf I}^{M/2} & {\bf 0}^{M/2} \\
{\bf 0}^{M/2} & {\bf 0}^{M/2}
\end{array}
\right]$}
\enq
and $\hat{S}_{M,i} S_{M,i}^\dagger S_{M,i} = S_{M,i}$.
Multiplying both sides of (\ref{nrx}) by
$U_{M,i}\hat{S}_{M,i}W_{M,i}^\dagger \ce_{M,i}^\dagger (H_n) = 
U_{M,i} \hat{S}_{M,i} S_{M,i}^\dagger U_{M,i}^\dagger$, we get
\beqa
\nonumber
U_{M,i} \hat{S}_{M,i} S_{M,i}^\dagger
U_{M,i}^\dagger \hat{X}_n({\bf s}) & = &
\sqrt{\frac{\rho}{M}} \ce_{M,i}(H_n) v_{M,i}({\bf s}) ~~~~~~~~ \\
\label{decode}
& & + U_{M,i} \hat{S}_{M,i} S_{M,i}^\dagger U_{M,i}^\dagger V_n,
\enqa
where we have used $\ce_{M,1}^\dagger(H_n) \ce_{M,2}(H_n) = 0$ to cancel
the contribution of other partition.  Note that using (\ref{ssmul}), it
follows that
$\hat{V}_n = U_{M,i} \hat{S}_{M,i} S_{M,i}^\dagger U_{M,i}^\dagger V_n$
has the same the statistics as $V_n$. We can rewrite
(\ref{decode}) as
\beq
\label{decode2}
\acute{X}_n({\bf s}) = \sqrt{\frac{\rho}{M}} S_{M,i} W_{M,i}^\dagger
v_{M,i}({\bf s}) + \hat{V}_n
\enq

Using (\ref{part}), one can iteratively generate the
equivalent channels for each partitions with
${\bf h}_{M,1} = [h_1, \cdots, h_M]$ and
${\bf h}_{M,2} = [h_{M+1}, \cdots, h_{2M}]$, as
\beq
\ce_{2M,1}(\beh) =
\left[
\begin{array}{rr}
\ce_{M,1}(\beh_{M,1}) & \ce_{M,2}(\beh_{M,2}) \\
\ce_{M,1}^*(\beh_{M,2}) & -\ce_{M,2}^*(\beh_{M,1})
\end{array}
\right],
\enq
\beq
\ce_{2M,2}(\beh) =
\left[
\begin{array}{rr}
-\ce_{M,2}(\beh_{M,1}) & -\ce_{M,1}(\beh_{M,2}) \\
-\ce_{M,2}^*(\beh_{M,2}) & \ce_{M,1}^*(\beh_{M,1})
\end{array}
\right]
\enq

\subsection{Codes with controllable decoding complexity}

Before we get to the code design, we first prove some properties
that are given in the following propositions.

\begin{prop}
\label{prop1}
The matrices
\beq
\bet_{2M,1}(\beh_{2M}) = \ce^{\dagger}_{2M,1}(\beh_{2M})\ce_{2M,1}(\beh_{2M}),
\enq
\beq
\label{tdef}
\bet_{2M,2}(\beh_{2M}) = \ce^{\dagger}_{2M,2}(\beh_{2M})\ce_{2M,2}(\beh_{2M}),
\enq
\beqa
\nonumber
\bek_{M}(\beh_{M,1},\beh_{M,2}) & = &
\ce^\dagger_{M,1}(\beh_{M,1})\ce_{M,2}(\beh_{M,2}) - \\
& & ~~~ \ce^{T}_{M,1}(\beh_{M,2})\ce_{M,2}^*(\beh_{M,1}),
\enqa
\beqa
\nonumber
\beu_{M}(\beh_{M,1},\beh_{M,2}) & = & \ce^{\dagger}_{M,1}(\beh_{M,1})\ce_{M,1}
(\beh_{M,2})+\\
& & \ce^\dagger_{M,1}(\beh_{M,2})\ce_{M,1}(\beh_{M,1}),
\enqa
\beqa
\nonumber
\bev_{M}(\beh_{M,1},\beh_{M,2}) & = & \ce^{\dagger}_{M,2}(\beh_{M,1})\ce_{M,2}
(\beh_{M,2})+\\
& & \ce^\dagger_{M,2}(\beh_{M,2})\ce_{M,2}(\beh_{M,1})
\enqa
are real $\forall$ $\beh_{2M} \in \compf^{2M\times 1}$, $\beh_{M,1} \in
\compf^{M \times 1}$, $\beh_{M,2} \in \compf^{M \times 1}$.
\end{prop}
\begin{proof}
Omitted.
\end{proof}

\ \\ \\
\begin{prop}
\label{propa2}
For any $\beh_{M,1}, \beh_{M,2} \in \compf^{M\times 1}$,
if $\beu_{M}(\beh_{M,1},\beh_{M,2})$ and $\bet_{M,1}(\beh_{M,1})$ have the same
eigenvectors and $\bev_{M}(\beh_{M,1},\beh_{M,2})$ and
$\bet_{M,2}(\beh_{M,1})$ have the same eigenvectors, then
for any $\beh_{2M}, \begn_{2M} \in \compf^{2M\times 1}$,
eigenvectors of $\beu_{2M}(\beh_{2M},\begn_{2M})$,
$\bet_{2M,1}(\beh_{2M})$ are the same, and similarly, the
eigenvectors of $\bev_{2M}(\beh_{2M},\begn_{2M})$,
$\bet_{2M,2}(\beh_{2M})$ are also the same.
\end{prop}
\begin{proof}
Omitted.
\end{proof}

\ \\ \\
\begin{prop}
\label{prop3}
If for any $\beh_{4M} \in \compf^{4M\times 1}$
\beq
\left[\begin{array}{rr}
\bea_{4M} \\
\beb_{4M}
\end{array}
\right]
\enq
is an eigenvector for $\bet_{4M,1}(\beh_{4M})$ with $\lambda_{4M}$ as
the associated eigenvalue, then the eigenvector of $\bet_{4M,2}(\beh_{4M})$
is
\beq
\left[\begin{array}{rr}
\beb_{4M} \\
-\bea_{4M}
\end{array}
\right]
\enq
with the same eigenvalue $\lambda_{4M}$.
Furthermore,
\beqa
\bek_{2M} \beb_{4M} & = & \lambda_{4M}^k \bea_{4M}, \\
\bek_{2M}^\dagger \bea_{4M} & = & \lambda_{4M}^k \beb_{4M},
\enqa
where the dependence of $\bek$ on the channel realization is dropped
for convenience. 
\end{prop}
\begin{proof}
Omitted.
\end{proof}

\ \\ \\
\begin{prop}
\label{prop4}
If
\beq
\left[\begin{array}{rr}
        \bea_{4M} \\
        \beb_{4M}
\end{array}\right]
\enq
is an eigenvector of $\bet_{4M,1}(\beh_{4M})$, then the eigenvectors of
$\bet_{8M,1}(\beh_{8M})$ are
\beq
{1 \over \sqrt{2}}
\left[\begin{array}{rr}
        \bea_{4M} \\
        \beb_{4M} \\
        \beb_{4M} \\
        -\bea_{4M}
\end{array}
\right]
~~ \mbox{and} ~~
{1 \over \sqrt{2}}
\left[\begin{array}{rr}
        \bea_{4M} \\
        \beb_{4M} \\
        -\beb_{4M} \\
        \bea_{4M}
\end{array}\right]
\enq
\end{prop}
\begin{proof}
We prove this by induction. It is easy to check it for $M=8$. Let us
assume that this is true for $\bet_{k,1}(\beh_{k})$ $\forall ~ k \leq 4M$
i.e. if $\left[\begin{array}{rr}
	\bea_{4M} \\
	\beb_{4M}
\end{array}\right]$ is an eigenvector of $\bet_{4M,1}(\beh_{4M})$,
then $\bea_{4M}=\left[\begin{array}{rr}
	\bea_{2M} \\
	\beb_{2M}
\end{array}\right]$
is an eigenvector of $\bet_{2M,1}(\beh_{2M})$.
By using Proposition \ref{prop1}, $\bet_{8M,1}(\beh_{8M})$ can be written as
into smaller parts as
\beq
\label{dum1}
\bet_{8M,1}(\beh_{8M}) =
\mbox {\small{$
\left[
\begin{array}{rr}
\bet_{4M,1}(\beh_{4M,1})+\bet_{4M,1}(\beh_{4M,2}) \hspace{10pt} \\
\bek_{4M}(\beh_{4M,1},\beh_{4M,2}) \\
\bek_{4M}^{T}(\beh_{4M,1},\beh_{4M,2}) \hspace{50pt} \\
\bet_{4M,2}(\beh_{4M,2})+\bet_{4M,2}(\beh_{4M,1})
\end{array}
\right]$}} 
\enq
We have to show that if
$\left[\begin{array}{rr}
	\bea_{4M} \\
	\beb_{4M}
\end{array}\right]$
is an eigenvector of $\bet_{4M,1}(\beh_{4M})$,
then
\beq
\label{dum2}
\bet_{8M,1}(\beh_{8M})\left[\begin{array}{rr}
	\bea_{4M} \\
	\beb_{4M} \\
	\beb_{4M} \\
	-\bea_{4M}
\end{array}\right] = \lambda_{8M}\left[\begin{array}{rr}
	\bea_{4M} \\
	\beb_{4M} \\
	\beb_{4M} \\
	-\bea_{4M}
\end{array}\right]
\enq
From the induction assumption, $\left[\begin{array}{rr}
	\bea_{4M} \\
	\beb_{4M} 
\end{array}\right]$ is an eigenvector of $\bet_{4M,1}(\beh_{4M,1})$ and
$\bet_{4M,1}(\beh_{4M,2})$ with eigenvalues $\lambda_{4M}$, $\lambda_{4M}^a$
respectively and using Proposition \ref{prop3}, these are also the
eigenvectors of $\bet_{4M,2}(\beh_{4M,1})$ and $\bet_{4M,2}(\beh_{4M,2})$.
Substituting in (\ref{dum1}) and (\ref{dum2}),  we have to show that
\beq
\label{dum3}
\bek_{4M}(\beh_{4M,1},\beh_{4M,2})\left[\begin{array}{rr}
	\beb_{4M} \\
	-\bea_{4M}
\end{array}\right]=\lambda_{8M}^a\left[\begin{array}{rr}
	\bea_{4M} \\
	\beb_{4M} \end{array}\right]
\enq
\beq
\label{dum4}
\bek_{4M}^\dagger (\beh_{4M,1},\beh_{4M,2})
\left[\begin{array}{rr}
	\bea_{4M} \\
	\beb_{4M} \end{array}\right]=\lambda_{8M}^a\left[\begin{array}{rr}
	\beb_{4M} \\
	-\bea_{4M}
\end{array}\right]
\enq
If $\left[\begin{array}{rr}
	\bea_{4M} \\
	\beb_{4M} \end{array}\right]$ is an eigenvector of
$\bet_{4M,1}(\beh_{4M})$, then it follows from
the induction assumption that $\bea_{4M}$ is an eigenvector of
$\bet_{2M,1}$ (and $\beu_{2M}$)
and $\beb_{4M}$ is an eigenvector of $\bet_{2M,2}$ (and $\bev_{2M}$),
where the dependence of $\bet$ on the channel realization is dropped
for convenience. Using Propositions \ref{propa2} and \ref{prop3}, we have
\beq
\label{dum5}
\bek_{2M}(\beh_{2M,1},\beh_{2M,2})\beb_{4M}=\lambda_{4M}^k
(\beh_{2M,1},\beh_{2M,2}) \bea_{4M},
\enq
\beq
\label{dum6}
\bek_{2M}^{\dagger}(\beh_{2M,1},\beh_{2M,2})\bea_{4M}=
\lambda_{4M}^{k}(\beh_{2M,1},\beh_{2M,2})\beb_{4M},
\enq
\beq
\label{dum7}
\beu_{2M}(\beh_{2M,1},\beh_{2M,2})\bea_{4M}=\lambda_{4M}^c
(\beh_{2M,1},\beh_{2M,2})\bea_{4M},
\enq
\beq
\label{dum8}
\bev_{2M}(\beh_{2M,1},\beh_{2M,2})\bea_{4M}=\lambda_{4M}^{c}
(\beh_{2M,1},\beh_{2M,2}) \beb_{4M}
\enq
Note that
{ {\scriptsize$\bek_{4M}(\beh_{4M,1},\beh_{4M,2})$
$=\left[
\begin{array}{rr}
-\bek_{2M}(\beh_{2M,1},\beh_{2M,3})+\bek_{2M}(\beh_{2M,4},\beh_{2M,2}) \hspace{30pt} \\ -\beu_{2M}(\beh_{2M,1},\beh_{2M,4})+\beu_{2M}^*(\beh_{2M,2},\beh_{2M,3}) \\
-\bev_{2M}(\beh_{2M,2},\beh_{2M,3})+\bev_{2M}^*(\beh_{2M,1},\beh_{2M,4}) \hspace{30pt} \\ \bek_{2M}^\dagger(\beh_{2M,1},\beh_{2M,3})-\bek_{2M}^\dagger(\beh_{2M,4},\beh_{2M,2})
\end{array}
\right],$
}}
hence using (\ref{dum5}), (\ref{dum6}), (\ref{dum7}), (\ref{dum8}),
we have
\beq
\bek_{4M}(\beh_{4M,1},\beh_{4M,2})\left[\begin{array}{rr}
	\beb_{4M} \\
	-\bea_{4M} \end{array}\right]=
\lambda_{8M}^a
\left[\begin{array}{rr}
\beb_{4M} \\
-\bea_{4M} \end{array}\right],
\enq
where $\lambda_{8M}^a = -\lambda_{4M}^k(\beh_{2M,1},\beh_{2M,3})
+ \lambda_{4M}^k(\beh_{2M,4},\beh_{2M,2})
+ \lambda_{4M}^c(\beh_{2M,1},\beh_{2M,4})
- \lambda_{4M}^c(\beh_{2M,2},\beh_{2M,3})$.
This proves (\ref{dum3}).
Hence
\[
\mbox{ {\scriptsize$\left[
\begin{array}{rr}
-\bek_{2M}(\beh_{2M,1},\beh_{2M,3})+\bek_{2M}(\beh_{2M,4},\beh_{2M,2}) \hspace{30pt} \\ -\beu_{2M}(\beh_{2M,1},\beh_{2M,4})+\beu_{2M}^*(\beh_{2M,2},\beh_{2M,3}) \\
-\bev_{2M}(\beh_{2M,2},\beh_{2M,3})+\bev_{2M}^*(\beh_{2M,1},\beh_{2M,4}) \hspace{30pt} \\ \bek_{2M}^\dagger(\beh_{2M,1},\beh_{2M,3})-\bek_{2M}^\dagger(\beh_{2M,4},\beh_{2M,2})
\end{array}
\right]$}}
\left[\begin{array}{rr}
\beb_{4M} \\
-\bea_{4M}
\end{array}\right]
\]
\begin{flushright}
$=\lambda_{8M}^{a}\left[\begin{array}{rr}
	\bea_{4M} \\
	\beb_{4M}
\end{array}\right] $
\end{flushright}
By interchanging the first half of the rows with the second half,
then interchanging the first half of the columns with the second half,
then multiplying the first half of columns and the second half of rows
with $-1$, and using the fact that $\beu_{M}$, $\bev_M$ are real,
Hermitian matrices, we can write the above equations as \\
{ {\scriptsize$ \left[
\begin{array}{rr}
-\bek_{2M}^\dagger(\beh_{2M,1},\beh_{2M,3})+\bek_{2M}^\dagger(\beh_{2M,4},\beh_{2M,2}) \hspace{30pt} \\ 
-\bev_{2M}^\dagger(\beh_{2M,2},\beh_{2M,3})+\bev_{2M}^{*^\dagger}(\beh_{2M,1},\beh_{2M,4}) \\
-\beu_{2M}^\dagger(\beh_{2M,1},\beh_{2M,4})+\beu_{2M}^{*^\dagger}(\beh_{2M,2},\beh_{2M,3}) \hspace{30pt} \\ \bek_{2M}(\beh_{2M,1},\beh_{2M,3})-\bek_{2M}(\beh_{2M,4},\beh_{2M,2})
\end{array}
\right]
$}}
$\left[\begin{array}{rr}
	\bea_{4M} \\
	\beb_{4M} \end{array}\right]$
\begin{flushright}
	$=\lambda_{8M}^{a}\left[\begin{array}{rr}
	\beb_{4M} \\
	-\bea_{4M}
\end{array}\right] $
\end{flushright}
or
\beq
\bek_{4M}^{\dagger}(\beh_{4M,1},\beh_{4M,2})\left[\begin{array}{rr}
	\bea_{4M} \\
	\beb_{4M} \end{array}\right]=\lambda_{8M}^a\left[\begin{array}{rr}
	\beb_{4M} \\
	-\bea_{4M} \end{array}\right]
\label{expla19}  
\enq
Hence if $\left[\begin{array}{rr}
	\bea_{4M} \\
	\beb_{4M}
\end{array}\right]$ is an eigenvector of $\bet_{4M,1}(\beh_{4M})$,
$\left[\begin{array}{rr}
	\bea_{4M} \\
	\beb_{4M} \\
	\beb_{4M} \\
	-\bea_{4M}
\end{array}\right]$
is an eigenvector of $\bet_{8M,1}(\beh_{8M})$.
Similarly, by using (\ref{dum3}) and (\ref{dum4}), one can show that
$\left[\begin{array}{rr}
	\bea_{4M} \\
	\beb_{4M} \\
	-\beb_{4M} \\
	\bea_{4M}
\end{array}\right]$ is also an eigenvector for $\bet_{8M,1}(\beh_{8M})$.
Q.E.D.
\end{proof}

\ \\ \\
\noindent \underline{\emph Example}: For $M=4$, the eigenvector matrix
for $T_{4,1}$ (or $W_{M,1}$ in (\ref{svdeq})) is computed as
\beq
W_{4,1} = {1 \over \sqrt{2}}
\left[\begin{array}{rr}
1 & 1 \\
1 & -1
\end{array}
\right]
\enq
and using Proposition \ref{prop4}, the eigenvector
matrix for $T_{8,1}$ is given by
\beq
W_{8,1} = {1 \over 2}
\left[\begin{array}{rrrr}
 1 &  1 &  1 &  1 \\
 1 &  1 & -1 & -1 \\
 1 & -1 & -1 &  1 \\
-1 &  1 & -1 &  1
\end{array}
\right]
\enq

\begin{figure}[ht]
\centering
\scalebox{0.55}{\input{cpxt_div.pstex_t}}
\caption{Diversity versus decoding complexity tradeoff for the proposed rate $1$
codes.}
\label{plot3}
\end{figure}
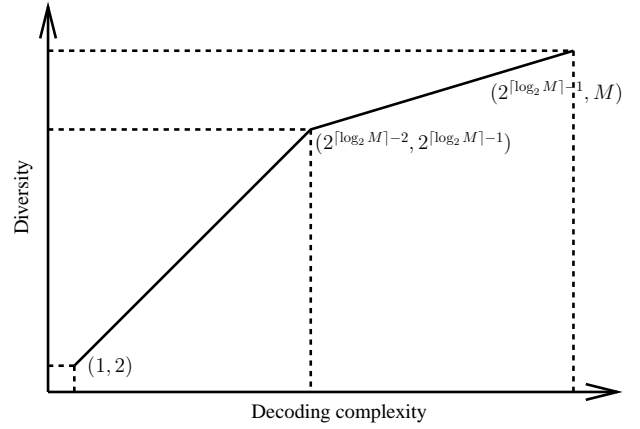

An important aspect of the eigenvector matrices of
$T_{M,1}$ and $T_{M,2}$ is that they are independent
of the channel realization.
This property is quite useful in constructing codes
with controllable ML decoding complexity.
\ \\
\begin{prop}
\label{prop5}
The eigenvalues for the first partition of the code matrix are given by
\beq
\left\{ Q_{M,1}^\dagger v_{M,1}({\bf s}),
Q_{M,1}^\dagger v_{M,1}({\bf s}^*) \right\}
\enq
where $Q_{M,1} = \sqrt{M/2} W_{M,1}$ and 
the determinant for the first partition of the code matrix is given by
\beq
\det\left\{G_{M}[\ca_{M,1}({\bf s})]\right\} =
f\left(Q_{M,1}^\dagger v_{M,1}({\bf s})\right)
\enq
and for any
$n$ length vector ${\bf q}$, $f({\bf q}) = \prod_{k=1}^n |q_i|^2$.
\end{prop}
\begin{proof}
Omitted.
\end{proof}
\ \\ \\
Result is similar for the second partition, and due to similarity
with the above Proposition, we omit it.

We note here from (\ref{decode2}) that it is $W_{M,i}$ that dictates
the ML decoding complexity. For example, we could precode the
information-carrying symbol vector ${\bf c}$ in (\ref{precod}) such that
\beq
v_{M,i} ({\bf c}) = W_{M,i} v_{M,i}({\bf s})
\enq
According to (\ref{decode2}), this code will admit single symbol
ML decoding. But this will give the determinant of the error code matrix as
\beq
\det\left\{G_{M}[\ca_{M,1}(W_{M,1}({\bf c} - {\bf e}))]\right\} =
{M \over 2} f \left( v_{M,i}({\bf c}-{\bf e}) \right)
\enq
Since the elements of ${\bf c}$ and ${\bf e}$ are drawn from the same
constellation, hence even if any element of ${\bf c}$ and ${\bf e}$
is the same, then
\beq
\min_{{\bf c},{\bf e},{\bf c} \neq {\bf e}}
\det\left\{G_{M}[\ca_{M,1}(W_{M,1}({\bf c} - {\bf e}))]\right\} = 0
\enq
For single symbol decoding, the minimum rank of $G_M$ would be $2$
(also the rank of $G_M^\dagger G_M$). In general, ML decoding of $M/2^n$,
$n = 1, 2, \cdots, \log_2M$,
symbols together would mean having the precoding matrix as a
block diagonal matrix with each block as $W_{M/2^{n-1},1}$ (scaled
appropriately) with constellation rotation to ensure that the
rank of $G_{M}[\ca_{M,1}({\bf c}-{\bf e})]$ is $M/2^{n-1}$.

The only way to achieve full diversity would be to choose
$n=1$ or decode $M/2$ information symbols together.
One can employ various methods like constellation rotation
given in \cite{tirkkonen2,np,xia,liu}.

\begin{figure}[t]
\centering
\includegraphics[height=2.1in]{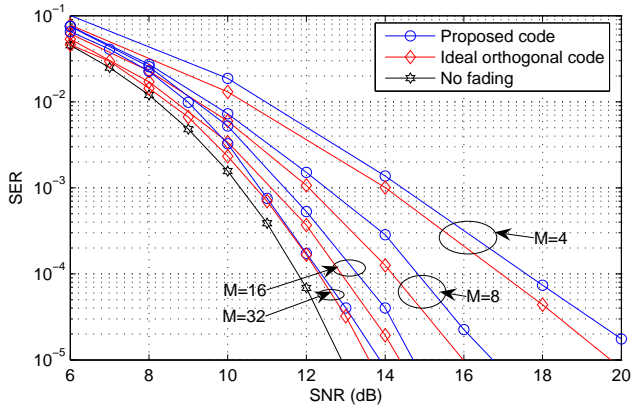}
\caption{SER versus SNR for various $M$ and $N=1$ with QPSK modulation
for the rate-$1$ constructed codes with $M/2$ symbols
decoded together and the \emph{ideal} orthogonal codes.}
\label{plot1}
\end{figure}

By using this block diagonal structure, we can construct
codes with ML decoding of different symbols together
and hence the ML decoding complexity can be controlled.
We plot the diversity versus complexity tradeoff in Fig. \ref{plot3},
where $M$ is not necessarily a power of $2$. The code design
for such $M$ is done by consturcting a code for $2^{\lceil \log_2M \rceil}$
transmit antennas that admits ML decoding complexity of
$2^{\lceil \log_2M \rceil-n}$ ($n=1,2,\cdots$)
and has rank of each partition as
$2^{\lceil \log_2M \rceil-n+1}$ and then deleting columns suitably chosen
to retain the same rank and to
have the code matrix of the size $2^{\lceil \log_2M \rceil} \times M$.

Note that it is not necessary to assume that 
$P = M$ i.e. unit rate. One could design codes with
$P < M$ that may have additional coding gain while
sacrificing code rate.


\section{Numerical Results}

The symbol error rate (SER) versus the average SNR per receive antenna for the
proposed rate-$1$ code that admits decoding in pairs of symbols is plotted in
Fig. \ref{plot1} with QPSK modulation for $M = 4,8,16$ and $N=1$. Also
plotted is the performance of an \emph{ideal} rate-$1$ orthogonal space-time
codes (non-existent for $M > 2$) with equivalent channel as $||H||_F$.

\begin{figure}[t]
\centering
\includegraphics[height=2.1in]{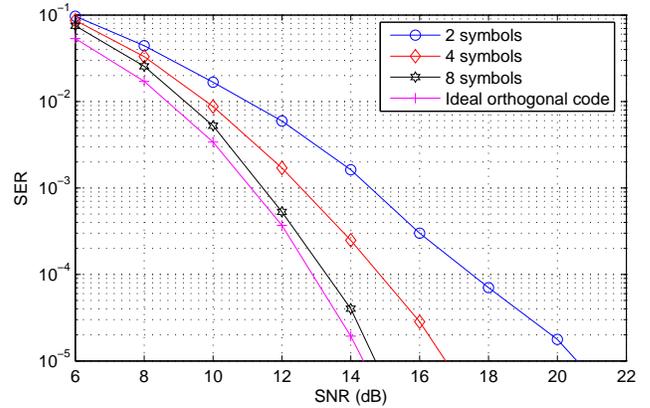}
\caption{SER versus SNR for $M=16$ and $N=1$ with QPSK modulation
for the rate-$1$ constructed and the \emph{ideal} codes for varying
decoding complexity.}
\label{plot2}
\end{figure}

Fig. \ref{plot2} plots the SER curves for $M=16$ and $N=1$ for different
ML decoding complexities. Fig. \ref{plot3} plots the diversity versus
complexity tradeoff for the proposed codes.

\section{Conclusions}

We have constructed a class of linear space-time codes that
have controllable ML decoding complexity for any number of transmit
antennas. The diversity versus decoding complexity tradeoff is shown.
We show that one can design rate $1$ codes that achieve performance
quite close to the rate $1$ ideal orthogonal codes (non-existent for
for $M>2$).

\end{document}

%% file: cpxt_div.pstex_t
\begin{picture}(0,0)%
\includegraphics{cpxt_div.pstex}%
\end{picture}%
\setlength{\unitlength}{3947sp}%
\begingroup\makeatletter\ifx\SetFigFont\undefined%
\gdef\SetFigFont#1#2#3#4#5{%
  \reset@font\fontsize{#1}{#2pt}%
  \fontfamily{#3}\fontseries{#4}\fontshape{#5}%
  \selectfont}%
\fi\endgroup%
\begin{picture}(7054,4915)(3780,-4343)
\put(4651,-3736){\makebox(0,0)[lb]{\smash{{\SetFigFont{14}{16.8}{\familydefault}{\mddefault}{\updefault}{$(1,2)$}%
}}}}
\put(7251,-1186){\makebox(0,0)[lb]{\smash{{\SetFigFont{14}{16.8}{\familydefault}{\mddefault}{\updefault}{$(2^{\lceil \log_2M \rceil - 2}, 2^{\lceil \log_2M \rceil -1})$}%
}}}}
\put(9251,-611){\makebox(0,0)[lb]{\smash{{\SetFigFont{14}{16.8}{\familydefault}{\mddefault}{\updefault}{$(2^{\lceil \log_2M \rceil -1},M)$}%
}}}}
\put(6526,-4261){\makebox(0,0)[lb]{\smash{{\SetFigFont{14}{16.8}{\familydefault}{\mddefault}{\updefault}{Decoding complexity}%
}}}}
\put(3976,-2161){\rotatebox{90.0}{\makebox(0,0)[lb]{\smash{{\SetFigFont{14}{16.8}{\familydefault}{\mddefault}{\updefault}{Diversity}%
}}}}}
\end{picture}%

%% file: isit.bbl
\begin{thebibliography}{9}


\bibitem{vahid}
V. Tarokh, N. Seshadri and A.R. Calderbank,
$``$Space-time codes for high data rate wireless communications : Performance
criterion and code construction,$"$
\emph{IEEE Trans. Inform. Theory}, vol. 44, pp. 744-765, March 1998.

\bibitem{block2}
V. Tarokh, H. Jafarkhani and A.R. Calderbank,
$``$Space-time block codes from orthogonal designs,$"$
\emph{IEEE Trans. Inform. Theory}, vol. 45, pp. 1456-1467, July 1999.

\bibitem{hott}
O. Tirkkonen and A. Hottinen,
$``$Square-matrix embeddable space-time block
codes for complex signal constellations,$"$
\emph{IEEE Trans. Inform. Theory}, vol. 48, pp. 384-395, Feb. 2002.

\bibitem{dasilva}
V.M. DaSilva and E.S. Sousa, $``$Fading-resistant modulation using
several transmitter antennas,$"$ \emph{IEEE Trans. Commun.}, vol.45,
pp. 1236-1244, Oct. 1997.

\bibitem{damen}
M.O. Damen, K. Abed-Meraim and J.-C. Belfiore, $``$Diagonal
algebraic space-time block codes,$"$ \emph{IEEE Trans. Inform. Theory},
vol. 48, pp. 628-636, March 2002.

\bibitem{rot2}
J. Boutros and E. Viterbo, $``$Signal space diversity: A power and
band-width efficient diversity technique for the Rayleigh fading
channel,$"$ \emph{IEEE Trans. Inform. Theory}, vol. 44, pp.
1453~V1467, July 1998.

\bibitem{jafar1}
H. Jafarkhani, $``$A quasi-orthogonal space-time block code,$"$
\emph{IEEE Trans. Commun.}, vol. 49, pp. 1-4, Jan. 2001.

\bibitem{tirkkonen1}
O. Tirkkonen, A. Boariu and A. Hottinen, $``$Minimal
non-orthogonality rate 1 space-time block code for 3+ Tx
antennas,$"$ in \emph{Proc. IEEE ISSSTA}, Parsippany, NJ, Sept. 2000.

\bibitem{tirkkonen2}
O. Tirkkonen, $``$Optimizing space-time block codes by
constellation rotations,$"$ in \emph{Proc. Finnish Wireless Commun.
Workshop 2001}, Oct. 2001.

\bibitem{pap_fos}
C. B. Papadias and G. J. Foschini, $``$Capacity-approaching space-time codes
for systems employing four transmitter antennas,$"$ \emph{IEEE Trans. on
Inform. Theory}, vol. 49, pp. 726-732, March 2003. 

\bibitem{np}
N. Sharma and C.B. Papadias, $``$Improved quasi-orthogonal codes through
constellation rotation,$"$
\emph{IEEE Trans. Commun.}, vol. 51, pp. 332-335, March 2003.

\bibitem{np2} N. Sharma and C.B. Papadias, $``$Full-rate full-diversity
linear quasi-orthogonal space-time codes for any number of transmit
antennas,$"$ \emph{Proc. Allerton Conf. Commun. Control Computing},
Monticello, IL, Oct. 2003, also in \emph{EURASIP J. Applied Signal
Processing (Special Issue on Advances in Smart Antennas)}, vol. 2004,
no. 9, pp. 1246-1256, Aug. 2004.

\bibitem{rajan1} Z. A. Khan and B. S. Rajan,
$``$Single-Symbol Maximum-Likelihood Decodable Linear STBCs,$"$
\emph{IEEE Trans. Inform. Theory}, vol.52, pp. 2062-2091, May 2006.

\bibitem{xia} D. Wang and X. Xia, $``$Optimal diversity product rotations
for quasi-orthogonal STBC with MPSK symbols,$"$ \emph{IEEE Commun. Lett.},
vol. 9, pp. 420-422, May 2005.

\bibitem{liu} L. Xian and H. Liu, $``$Optimal rotation angles for
quasi-orthogonal
space-time codes with PSK modulation,$"$ \emph{IEEE Commun. Lett.}, vol. 9,
pp. 676-678, Aug. 2005.

\end{thebibliography}
